# Quantum detector of noise based on a system of asymmetric Al superconducting rings

V.L. Gurtovoi, S.V. Dubonos, R. V. Kholin, A.V. Nikulov, N.N. Osipov, and V.A. Tulin
Institute of Microelectronics Technology, Russian Academy of Sciences, 142432 Chernogolovka, Moscow District, Russia

The future realization of quantum computers is closely connected with solid-state quantum bit (qubit) and coupled qubit fabrication and investigation of their functionality. The most promising candidates for solid-state qubits are two-level superconducting structures which act like artificial atoms with quantized energy levels (1, 2). To carry out quantum computations, decoherence time for superposition of macroscopic quantum states in superconducting structures should be longer than computation time. Decoherence is determined by the equilibrium and nonequilibrium noise environment, which have to be investigated and strictly controlled. Although, the intensity of equilibrium thermal fluctuations vanishes linearly with temperature, nevertheless, there are quantum fluctuations due to zero-point motion even at zero temperature. To detect nonequilibrium and vanishing with temperature equilibrium thermal fluctuations and separate them from quantum fluctuations, one should use a *quantum system*, i.e. a quantum detector of noise (QDN).

. As a basic element of the QDN, we propose to use an asymmetric superconducting ring (ASR) with or without Josephson junctions. Analogous to SQUID structures in magnetic field (3), ASR structures are able to rectify applied ac current or both external and internal noise without bias current when narrow part of the ASR is in the resistive state induced by the magnetic flux quantization persistent current. In this work, we have investigated magnetic field, temperature, and ac bias current dependence of the sensitivity for the ASR structures as QDsN. Calibration of the noise detector and optimization of the sensitivity has been carried out. Connected in series 18 ASRs have been investigated for the sensitivity improvement.

Investigated test structures (without Josephson junctions) consisted of asymmetric aluminum rings (45-50 nm thick, thermally evaporated on oxidized Si substrates) with semi-ring width of 200 and 400 nm for narrow and wide parts, respectively. 4 µm diameter single ASR and 20 ASR structures were fabricated by e-beam lithography and lift-off process. Fig. 1 shows an SEM image of the 20 ring structure. For this structures, the sheet resistance was 0.23 Ω/□ at 4.2 K, the resistance ratio R(300 K)/R(4.2 K)=2.7, and superconducting transition temperature was 1.24-1.27 K. Estimated coherence length $\xi$(T=0 K) was 170 nm and penetration depth $\lambda$(T=0 K) was 80 nm.

Measurements were carried out by applying 5 or 40 kHz sinusoidal bias current to current leads, whereas, rectified dc signal was measured in a frequency band from 0 to 30 Hz by home made preamplifier (followed by low-noise preamplifier SR560) at potential leads. Noise level of the amplification system was 20 nVpp for $f_b$=0 to1 Hz. It should be noted that rectification effects do not depend on frequency of the bias current at least up to 1 MHz [4]. Magnetic field direction was perpendicular to the ring's plane. Magnetic field time scanning was slow enough so that upper frequency of signal spectrum, which resulted from magnetic field changes, did not exceed 30 Hz. All signals corresponding to rectified voltage, current, and magnetic field were digitized by a multichannel 16-bit analog-digital converter card.

To make calibration of the QDN, ac bias current with different amplitude was applied to substitute noise with different amplitude level. Fig.2 shows typical oscillations of the rectified voltage as a function of magnetic field at different ac bias currents for the 18 ASRs in series (T = 1.244 K=0.98$T_c$). The oscillations are periodic with period equals to the superconducting flux quantum $\Phi_o$=h/2e, the positive and negative maxima being observed at $\Phi \approx \Phi_o(n\pm1/4)$, (n=0,±1,…) showing points of magnetic field with maximum rectification. Our experiments have shown that the rectified voltage for N ASRs is approximately N times higher compared to that for a single ASR fabricated in the same technological process. Therefore, the following qualitative explanations are given for the single ASR. Both AC bias current and relatively small in amplitude (due to weak screening) circular persistent current (which is due to magnetic flux quantization when the sum of the externally applied flux and flux induced by superconducting screening current are quantized) are responsible for the formation of the oscillations. Slowly varying in time, persistent current breaks the clockwise-anticlockwise symmetry of the currents flowing in the ring, which results in asymmetry of current-voltage characteristics. When persistent current flows in a definite direction, the bias current is added during its one half period to persistent current or subtracted during the opposite half period from that in a narrow part of the ring and vice versa in a wide part. When amplitude of the bias current is increased, first the criticality is achieved in the narrow part of the ring creating a voltage pulse of a proper sign, which after integration for many periods (frequency of the bias current is much higher than that of persistent current) forms rectified signal. Further increase in the bias current amplitude results in achieving of criticality in the wide part of the ring and formation (half-period shifted with respect to the narrow part) of an opposite sign voltage pulse. Time averaging of the voltage pulses of different polarity from narrow and wide parts of the ASR is the reason of rectified voltage diminishing when the amplitude of the bias current is increased. For the opposite direction of persistent current, there would be generated rectified signal of the opposite sign, which explains the sign-varying oscillations of the rectified voltage in magnetic field (Fig.2). From this, one can make conclusion that rectified dc voltage oscillations in magnetic field are proportional to persistent current, whereas ac bias current (noise) only creates conditions for observation of the oscillations, i.e. shifts the total current of the ASR to the critical value. The given above qualitative explanation of the rectified voltage behavior with increasing of the bias current is supported by experimental results in Fig.3, where the amplitude of the rectified voltage oscillations at $\Phi \approx \Phi_0/4$ is plotted as a function of the bias current amplitude. This dependence is nonmonotonous with sharp peak at near the critical current value. At low bias currents, the rectified voltage is zero since the ASR is in superconducting state.

To determine the exact position of the maximum in Fig.3 with respect to the critical current value, in Fig.4 we present temperature dependence of both the critical current of the single ASR obtained from current-voltage characteristics and the amplitude of the bias current $I_{max}$ at which the maximum of the rectified oscillations is observed. As clearly seen from Fig.4, the maximum of the oscillations is achieved when the bias current amplitude is slightly higher than the critical current of the ASR, having practically the same temperature dependence. Temperature dependence of the maximum rectified voltage $V_{max}$ is also shown in Fig.4. The maximum voltage increases with temperature decrease in a way similar to the bias current $I_{max}$, having linear dependence on the bias current (not shown hear). To the accuracy of measurement errors, the ratio $V_{max}/I_{max}=1.3$ Ohm is independent of temperature. When the amplitude of the bias current $I_0$ is equal to $I_{max}$, this ratio determines the maximum rectification efficiency which is equal to $V_{max}/I_0R_n=V_{max}/I_{max}R_n\approx 0.24$, where $R_n=5.5$ Ohm is the resistance of the ASR in the normal state. Since $I_{max}\approx I_c$, the maximum rectification efficiency could be increased using weak links or Josephson junctions which diminish the critical current (rf-SQUID or asymmetric dc-SQUID structure).

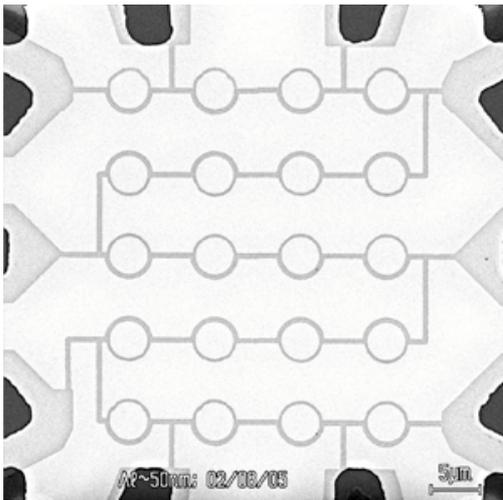

**Figure 1:** An SEM image of a superconducting structure with 20 asymmetric aluminium rings.

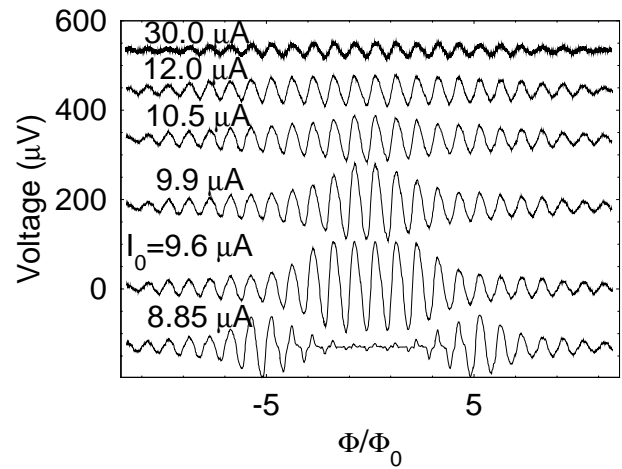

**Figure 2:** The quantum oscillations of the rectified dc voltage on 18 rings in magnetic field for different ac bias currents (f=40 kHz and amplitudes $I_0$=8.85; 9.6; 9.9; 10.5; 12; 30 μA) at T=1.244 K=0.98$T_c$. Except for $I_0$ = 9.6 μA, all curves are vertically shifted.

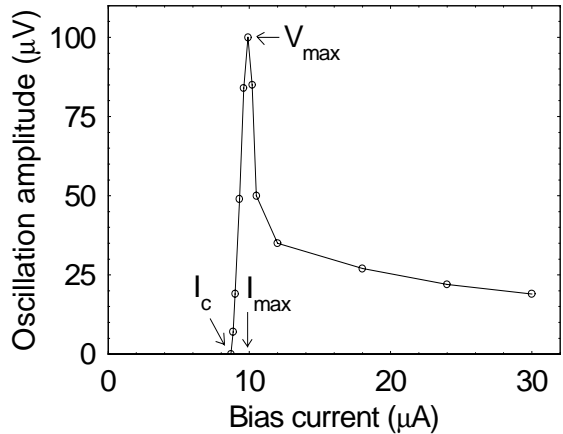 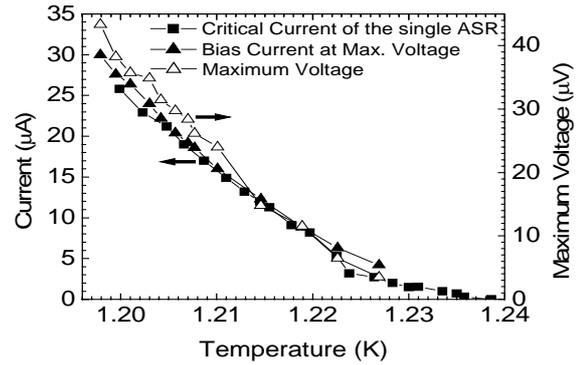

**Figure 3:** AC bias current dependence of the oscillation amplitude ($\Phi \approx \Phi_0/4$) for 18 rings at T= .244 K$\approx 0.98 T_c$.

**Figure 4:** Temperature dependence of the critical current $I_c$ of the single ASR, the amplitude of bias current $I_{max}$ corresponding to the maximum of oscillation amplitude, and the maximum voltage $V_{max}$ of the oscillation amplitude (for the explanation of $I_c$, $I_{max}$, and $V_{max}$ see Fig.3).

Our preamplifier noise level allows reliable measurement of 30 nV voltage signal. This allows to measure the current noise $I_{max}=V_{max}/(0.24R_n)\approx 20$ nA at different temperatures $T/T_c$. By fitting the critical current temperature dependence in Fig.4, one can get that $I_{max}\approx I_c=4.4$ mA$\times (1 - T/T_c)^{3/2}$. Since the bias current or noise only shift the QDN total current to near critical value $I_{max}$ where it has the maximum rectification efficiency, measurements of noise at different temperatures near $T_c$ allows amplitude profiling of the noise pulses in accordance with $I_c$ temperature dependence. Thus, the noise with the lower level will be detected by the QDN without biasing at temperature nearer to $T_c$. In case of high level of intrinsic noise in the amplification system, the QDN with N ASRs could overcome this lack of the amplifier sensitivity. Finally, let us estimate the level of the equilibrium noise. The root mean square of the equilibrium Nyquist noise can be determined from $R\langle I_{Ny}^2\rangle = k_B T\Delta\omega$ for the frequency range from 0 to the quantum limit $k_B T/\hbar$, i.e. $\langle I_{Ny}^2\rangle^{1/2}= k_B T/(\hbar R)^{1/2}$ which is equal to ~0.4 µA at $T\approx 1$ K and $R\approx 10$ Ohm. Thus, even the QDN based on a single ASR is able to detect temperature dependent equilibrium thermal fluctuations and separate them from temperature independent quantum fluctuations.

This work has been supported by a grant of the Program "Low-Dimensional Quantum Structures", the Russian Foundation of Basic Research (Grant 04-02-17068) and a grant of the program "Technology Basis of New Computing Methods".